\documentclass{article}

\usepackage{PRIMEarxiv}

\usepackage[utf8]{inputenc} % allow utf-8 input
\usepackage[T1]{fontenc}    % use 8-bit T1 fonts
\usepackage{hyperref}       % hyperlinks
\usepackage{url}            % simple URL typesetting
\usepackage{booktabs}       % professional-quality tables
\usepackage{amsfonts}       % blackboard math symbols
\usepackage{microtype}      % microtypography
\usepackage{fancyhdr}       % header
\usepackage{graphicx}       % graphics

\title{The Curvature of Spectral Energy Distribution and $\gamma$-ray Dominance of Fermi BL Lac Objects}

\author
{ANJUM Muhammad Shahzad$^{1,2}$,\quad CHEN Liang$^{1}$,\quad GU Minfeng$^{1}$\\
\\
\\
1. Key Laboratory for Research in Galaxies and Cosmology,\\
Shanghai Astronomical Observatory, Chinese Academy of Sciences\\
80 Nandan Road, Shanghai 200030, China\\
\\
2. University of Chinese Academy of Sciences\\
19A Yuquan Road, Beijing 100049, China\\
\\
\texttt{mshahzadanjum@yahoo.com}
}

\begin{document}
\maketitle

\begin{abstract}
The extragalactic $\gamma$-ray sky is dominated by blazars and their study plays an important role in understanding jet physics, cosmic evolution history and origin of ultra high energy cosmic rays. In this work, we study a large sample of BL Lac objects to investigate why some sources are detected in $\gamma$-rays, while others not. We selected 170 BL Lac objects, with measured synchrotron spectral curvature and Doppler factor, and divided them into Fermi-LAT detected (FBLs) and non-detected (NFBLs) sources. We show that FBLs have smaller curvature than NFBLs, even after getting rid of Doppler beaming effect. The BL Lac objects PKS 0048-09 and S5 0716+714 have similar synchrotron peak frequency and luminosity but different $\gamma$-ray dominance and their quasi-simultaneous broadband spectral energy distributions (SEDs) can be well fitted by a log-parabolic synchrotron self-Compton (SSC) model with same jet parameters except for the curvature and source size, assuming curvature being proportional to the size of emission region. Our results imply that for a given synchrotron luminosity, the different SED curvature and Compton dominance may account for the discrepancy between FBLs and NFBLs. We discuss these results in context of stochastic particle acceleration and radiation mechanisms.
\end{abstract}


\section{Introduction} \label{sec:introduction}
Beginning from the first identification of extragalactic source 3C 273 \cite{1978Natur.275..298S}, $\gamma$-ray astronomy has made great progress due to unprecedented detection capability of the new generation of telescopes (see, for example, \cite{2016ARA&A..54..725M} for recent reviews). The large area telescope (LAT) onboard Fermi satellite has detected more than 3000 $\gamma$-ray sources after first four years of operation, out of which about half are extragalactic sources \cite{2015ApJS..218...23A}. The extragalactic $\gamma$-ray sky is dominated by emission from blazars ($\sim 98\%$) \cite{2015ApJ...810...14A}, which is an extreme subclass of active galactic nuclei (AGN) with relativistic jets pointing close to our line of sight. Blazars have been classified as BL Lac objects (BL Lacs) or FSRQs according to whether the rest-frame equivalent width of their broad emission lines was greater than (for FSRQs) or smaller than (for BL Lacs) 5 {\AA}. Due to very high luminosity and violent variability, their $\gamma$-ray emission is believed to be produced in the relativistic jets making a small angle to our line of sight. The intensity is significantly boosted by the Doppler beaming effect because of light aberration through transformations $I_{\nu}=\delta^{3} I_{\nu'}'$ with $\delta=\nu/\nu'$ from intrinsic to observational frames, with the primed quantities referring to values measured in jet co-moving frame, while $\delta=1/[\Gamma(1-\beta\cos\theta)]$ is the Doppler beaming factor.

%% Gamma rays
The study of $\gamma$-ray emission of blazars plays an important role in understanding the jet physics (jet launching, energy dissipation and particle acceleration), the cosmic evolution history and the origin of ultra high energy cosmic rays \cite{1995Natur.374..430P, 2013PhRvL.111d1103K, 2016ARA&A..54..725M, 2015A&ARv..24....2M}. The $\gamma$-rays can be explained as inverse Compton (IC) emission of accelerated electrons in the jet. The IC emission in BL Lacs arises due to internal synchrotron photons and called synchrotron self-Compton (SSC) \cite{2011ApJ...736..131A}. The multi-wavelength catalog of blazars, Roma-BZCAT\footnote{http://www.asdc.asi.it/bzcat/} v5.0, is the most comprehensive catalog in the literature, which contains 3561 confirmed blazars or candidates \cite{2009A&A...495..691M, 2015Ap&SS.357...75M}.

Recently, Fermi Fourth LAT AGN Catalog (4LAC) presented a sample of nearly 3000 blazars or blazar candidates at high galactic latitude \cite{2020ApJ...892..105A}. Therefore, there may be $>$500 blazars having not been detected with $\gamma$-ray emission ($\gamma$-ray quiet), although they have similar emission properties at lower energy wavelength bands (radio through X-ray) compared to $\gamma$-ray detected blazars. The origin of this discrepancy is not yet known. Many studies have focused on this question and found that $\gamma$-ray loud blazars have relatively large jet opening angle, larger apparent superluminal velocity and higher brightness temperature relative to $\gamma$-ray quiet blazars \cite{2012ApJ...744..177L, 2012A&A...544A..34P, 2012ApJ...758...84P, 2014A&A...562A..64W, 2016AJ....152...12L}. All these properties are attributed to a relativistic geometric effect - Doppler beaming effect . The smaller viewing angle in $\gamma$-ray loud blazars makes apparent larger jet opening angle and larger Doppler beaming factor, which give rise to larger apparent superluminal velocity and higher brightness temperature. In addition to this geometric effect, it is expected that the origin of the discrepancy between $\gamma$-ray loud versus $\gamma$-ray quiet blazars might be intrinsic. However, it is not yet known.

%% SED and motivation
The spectral energy distribution (SED) of synchrotron emission from blazars is usually curved and presents a significant bump in $\nu-\nu f_{\nu}$ frame, which peaks between infrared to X-ray bands \cite{2011ApJ...735..108C, 2014Natur.515..376G, 2015MNRAS.448.1060G}. The curvature of broadband SED of the synchrotron bump is different among various blazars, which means that some SED bumps are broad while others are narrow \cite{2014ApJ...788..179C, 2018ApJS..235...39C}. If the SED (broad band or single band) is fitted by a log-parabolic function, $\log\nu f_{\nu}=\log\nu_{\rm p}f_{\nu_{\rm p}}-b(\log\nu-\log\nu_{\rm p})^{2}$, the coefficient of the second order term $b$ can be considered a "surrogate" to measure the curvature. The curvature is found to be negatively correlated with the SED peak frequency, which offers supporting evidence of a statistical or stochastic particle acceleration at work in blazar jets, through studying the broad band observations \cite{2014ApJ...788..179C} or single band (X-ray) observations \cite{2004A&A...413..489M}. It should be noted that the broadband curvature of a blazar is not changed with the transformation from observer to AGN frame (Cosmological redshift) and jet co-moving frame (Doppler beaming). However the single band curvature is effected due to Doppler effect causing the spectral shifts. In this work, we investigate the possible relation between $\gamma$-ray emission and the SED curvature for BL Lacs. The curvature in SED is produced due to an intrinsic curvature in the emitting electron energy distribution (EED) arising due to competition of particle acceleration and cooling \cite{1962SvA.....6..317K, 2011ApJ...739...66T}. The Section \ref{sec:sample} describes our sample and Section \ref{sec:results} presents the methods and a discussion on our results. We summarize our findings in Section \ref{sec:summary}. We assume a $\Lambda$CDM cosmology with values from the Planck results in our calculation; in particular, $\Omega_{m}=0.32$, $\Omega_{\Lambda}=0.68$, and the Hubble constant $H_{0}=67$ km s$^{-1}$ Mpc$^{-1}$ \cite{2014A&A...571A..16P}.

\section{The Sample} \label{sec:sample}
Starting from Mrk 421 \cite{2004A&A...413..489M}, the \cite{2008A&A...478..395M} found that curvature is an important feature of BL Lacs in X-rays. Even the Fermi-LAT $\gamma$-ray spectra of bright blazars show curvature \cite{2015ApJ...810...14A}. The high energy photon spectra can be fitted by a log-parabolic law \cite{2004A&A...413..489M, 2004A&A...422..103M},
\begin{equation}
F(E)= K \left(\frac{E}{E_0}\right)^{-\alpha -\beta \log(E/E_0)},
\end{equation}
with spectral index $\alpha$ at reference energy $E_0$ and $\beta$ is the curvature of bump. The corresponding peak energy of the spectral bump becomes
\begin{equation}
E_p = E_0 10^{\left( \frac{2-\alpha}{2\beta} \right)}.
\end{equation} 
\cite{2015ApJ...810...14A} found that the LAT spectra of $\sim$160 blazars show deviation from a power-law $F(E)=KE^{-p}$ and fitted with log-parabolic law. This suggests that the curvature is a common feature of blazar SED in any energy band.

The Fermi-LAT detected 1591 blazars at high galactic latitude in its Third LAT AGN Catalog (3LAC) based on first 4 years of operations \cite{2015ApJ...810...14A}. Based on 3LAC, \cite{2016ApJS..226...20F} collected available spectral data of 1392 blazars from NED\footnote{http://ned.ipac.caltech.edu/} to build broadband SEDs of synchrotron bump and fitted them by a log-parabolic function. We compile the synchrotron spectral curvature ($b$), synchrotron peak flux ($F_{\rm s}=\nu_{\rm p}f_{\nu_{\rm p}}$), and integrated $\gamma$-ray flux $F_{\gamma}$ in 0.1-100 GeV band for confirmed 620 BL Lacs in 3LAC \cite{2016ApJS..226...20F}. \cite{2007A&A...466...63W} presented the Doppler factors of a BL Lacs. Cross matching two catalogs, we compile the spectral parameters and Doppler factors for a sample of total 170 BL Lacs.

\section{Methods and Results} \label{sec:results}
A curved SED necessary suggests that the underlying emitting EED may be curved. A log-parabolic EED can arise due to stochastic nature of acceleration gain. \cite{2020ApJ...898...48A} found that only BL Lacs show the signature of stochastic acceleration, whereas FSRQs do not show any signature of such acceleration mechanism. The broadband SED in the log-parabolic model can be described as
\begin{equation}
\log\nu f_{\nu}=\log\nu_{\rm p}f_{\nu_{\rm p}}-b(\log\nu-\log\nu_{\rm p})^{2},
\end{equation}
where $b$ measures the width (curvature) of the SED bump. The synchrotron bump is particularly important and reveals the intrinsic curvature of EED. Since the accretion disk continuum emission in FSRQs are usually presented as a big blue bump at ultraviolet wavelength band, affecting accurate measurement of the curvature parameter, we only selected BL Lacs for purpose of our study. The selection of BL Lacs is further motivated by the fact that the high energy bump in FSRQs arises due to external Compton (EC) emission and, therefore, their curvature of IC bump may be related to the complex seed photon distribution and any absorption \cite{2014ApJ...788..179C}. Although the curvature of high energy bump in BL Lacs mimics the synchrotron curvature as it arises due to internal synchrotron seed photons \cite{2006A&A...448..861M}, it is hard to constrain the intrinsic shape of the EED from the IC bump, since the observed SED curvature of IC bump may also depend on the IC cooling regime. \cite{2009A&A...504..821P} found that the IC curvature in Thomson regime is smaller than in Klein-Nishina regime. Therefore, we constrain the intrinsic curvature of EED $r$ from synchrotron spectral curvature $b$ rather $\gamma$-ray curvature $\beta$. The intrinsic curvature of EED $r\sim 5b$ \cite{2006A&A...448..861M} and $\gamma$-ray emission in BL Lacs is expected to be intrinsically related and can be used to investigate the origin of discrepancy between Fermi detected BL Lacs (FBLs) and non-detected ones (NFBLs).

\subsection{FBLs versus NFBLs}
A large amount of multi-wavelength data for the objects in the Mets\"{a}hovi radio observatory BL Lac sample \cite{2006A&A...445..441N} was collected by \cite{2006A&A...445..441N}, which is supposed to have no selection criteria (other than declination) in addition to the ones in the original surveys \cite{2006A&A...445..441N}. The Mets\"{a}hovi radio observatory BL Lac sample includes 381 objects selected from the Veron-Cetty \& Veron BL Lac Catalog \cite{2000cqan.book.....V}, and 17 objects from the literature, of which many sources are from the well-known BL Lac samples like 1Jy, S4, S5, Einstein Medium Sensitivity Survey (EMSS), Einstein Slew Survey, and ROSAT Deep X-Ray Radio Blazar Survey (DXRBS). Based on the multi-wavelength data, the SED of each source were constructed in the $\log\nu-\log\nu F_{\nu}$ representation \cite{2006A&A...445..441N}, of which the synchrotron bump was fitted with a log-parabolic function.

Based on this sample, \cite{2007A&A...466...63W} collected the available data at 330 MHz, 360 MHz, 408 MHz, and 1.4 GHz from the Astrophysical Catalogs Support System (CATs) maintained by the Special Astrophysical Observatory, Russia, and also the available VLA or MERLIN core and extended flux, resulting in a sample of 170 BL Lacs. The low frequency radio power can be a reliable indicator of the intrinsic radio power, while the Doppler beaming can affect the observed radio power of the core \cite{2001ApJ...552..508G, 2004ApJ...613..752G}. Therefore, with the VLA or MERLIN core and the 408 MHz luminosity and assuming a jet speed $\Gamma=5$, consistent with Mets\"{a}hovi radio monitoring studies \cite{1999ApJ...521..493L}, the Doppler beaming factors have been estimated \cite{2007A&A...466...63W}. We compiled the Doppler factors and curvature parameters of these 170 BL Lacs. By cross correlating this sample with 3 LAC \cite{2015ApJ...810...14A}, 34 out of these 170 BL Lacs are not found to have $\gamma$-ray emission (NFBL). These NFBLs are: NRAO 5, NPM1G +41.0022, 1ES 0145+138, MS 0158.5+0019, MS 0257.9+3429, RXS J0314.0+2445, S5 0454+84, MS 0607.9+7108, B3 0651+428, 4C 22.21, RXS J0916.8+5238, B2 0927+35, RGB J0952+656, 1ES 1044+549, 1ES 1212+078, 1ES 1255+244, 1ES 1320+084N, MC 1400+162, MS 1407.9+5954, RGB J1427+541, MS 1443.5+6349, RXS J1516.7+2918, MS 1534.2+0148, RXS J1602.2+3050, RXS J1644.2+4546, RGB J1652+403, B3 1746+470, RXS J1750.0+4700, RGB J1811+442, PKS 2254+074, Q J2319+161, 1ES 2326+174, MS 2336.5+0517 and MS 2347.4+1924. The remaining 136 BL Lacs are, therefore, considered FBLs. Although many of these NFBLs might be detected in the 4LAC, however their $\gamma$-ray flux must be lower than the FBLs.
\begin{figure}
\centering
\includegraphics[height=6cm, width=8cm]{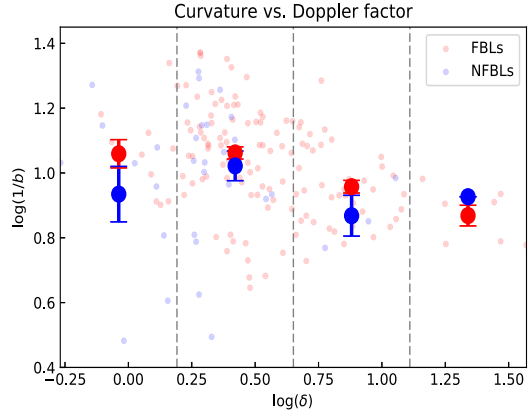}
\caption{The curvature and Doppler beaming effect in BL Lacs. The dashed lines represent 4 bins of Doppler factor $\delta$. The small points represent the values of individual sources, while the large points show the average value of curvature in each bin. The error bars correspond to one standard deviation of the average value.}
\label{fig:FBL_NFBL0}
\end{figure}

The SED of blazars is usually curved even in a single energy band. The Doppler factor changes cause the peak $\nu_p= \nu_p' \delta$ shifts in the SED ($\nu I_\nu =\delta^4 \nu' I_\nu'$) which affects the spectral curvature of blazars in Fermi-LAT $\gamma$-ray band. Other than the peak shifts, the intrinsic difference of curvature may account for discrepancy between FBLs and NFBLs. We reduce the effect of the relativistic Doppler beaming and directly compare FBLs and NFBLs. We searched the literature for published and archival observations that allow us obtain both curvature parameters and Doppler beaming factors of our Fermi BL Lacs. We plot curvature parameter (in $1/b$) against $\delta$ for FBLs and NFBLs in Figure \ref{fig:FBL_NFBL0}, which shows that both FBLs and NFBLs overlap but the curvature of FBLs would be on average smaller than that of NFBLs. We binned the data in $\log(\delta)$ and the find the average value of curvature in each bin, represented by big circles in Figure \ref{fig:FBL_NFBL0}. Due to only one NFBL in the last bin, we consider it as the average curvature value. FBLs, on average, seem to have smaller curvature (higher value of $1/b$) and, thus, broader SED than NFBLs.

\begin{figure}
\centering
\includegraphics[height=5cm, width=7cm]{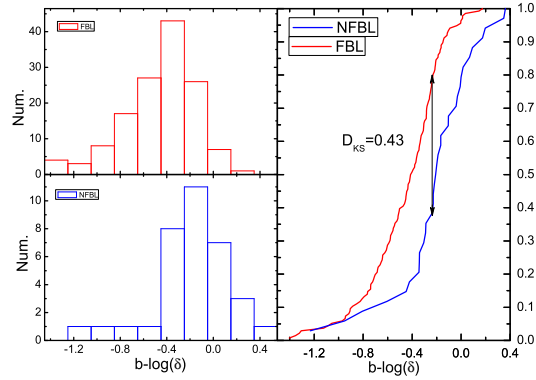}
\caption{The distribution and the cumulative fractions of FBLs versus NFBLs. The Kolmogorov-Smirnov (KS) test yields the significance level probability for the null hypothesis that FBLs and NFBLs are drawn from the same distribution $P=5.87\times10^{-5}$ and the statistic (the maximum separation of the two cumulative fractions) $D_{\rm KS}=0.43$.}
\label{fig:FBL_NFBL}
\end{figure}
In order to eliminate effect of Doppler beaming as much as possible, we artificially defined a new parameter, $b-\log\delta$, which roughly measures the curvature given the same beaming factor. Figure \ref{fig:FBL_NFBL} shows the distributions of $b-\log\delta$ and cumulative fraction of FBLs and NFBLs. It can be seen that statistically FBLs have relative smaller values of $b-\log\delta$ than that of NFBLs, which confirm the above finding that smaller curvature of synchrotron SED bump is attributed to larger $\gamma$-ray power. The Kolmogorov-Smirnov (KS) test yields the significance level probability for the null hypothesis that FBLs and NFBLs are drawn from the same distribution $P=5.87\times10^{-5}$, and the maximum separation of the two cumulative fractions is $D_{\rm KS}=0.43$. The TeV BL Lacs (TBLs) with synchrotron peak frequency $\nu_{p}>10^{15}$ Hz showed a similar behavior \cite{2011ApJ...739...73M}. \cite{2011ApJ...739...73M} and \cite{2013ApJS..207...16M} found X-ray band curvature of TBLs to be systematically smaller than that of non-detected at TeV energies. They suggested that X-ray flux can be used as a predictor for TeV $\gamma$-ray detection of BL Lacs. However the TeV spectral curvature may not correspond to X-ray curvature, as the TeV emission is significantly attenuated by extraglactic background light (EBL). Since GeV $\gamma$-ray emission is not absorbed, we argue that it can provide better constraint on TeV detection as compared to X-rays.

\subsection{SED Curvature and Gamma-ray Dominance}
\begin{figure}
\centering
\includegraphics[height=5cm, width=7cm]{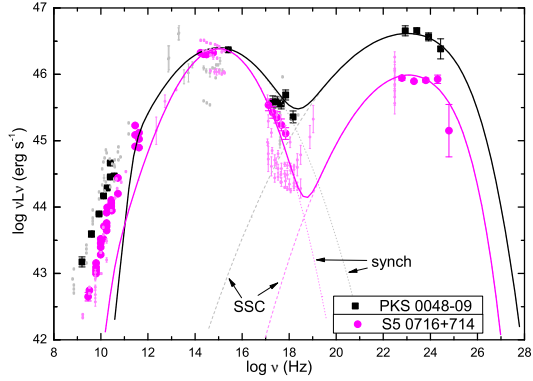}
\caption{Broadband SEDs of PKS 0048-09 and S5 0716+714. Quasi-simultaneous data points are presented as large squares (PKS 0048-09) and circles (S5 0716+714), while the small ones show the non-simultaneous data. The solid lines represent the synchrotron+SSC model curves. Both sources have same jet parameters other than size $R$ and curvature $r$ assuming $r\propto R$.}
\label{fig:theory}
\end{figure}
The fact that synchrotron SED is significantly curved inevitably implies a curvature in the steady electron energy distribution $N(\gamma)$. This curvature may be a result of particle cooling and acceleration which might be relevant on different times. Based on scenario of re-acceleration rather than continuous injection, the curvature can be related to stochastic acceleration term in the Fokker-Planck kinetic equation \cite{1962SvA.....6..317K, 2004A&A...413..489M}. A more efficient acceleration of stochastic type makes energy distribution electrons relatively broader, manifested by a smaller curvature \cite{2011ApJ...739...66T}. A higher acceleration efficiency implies a lesser time ($t=R/c$) spent in acceleration region by the particles and large number of acceleration steps, which corresponds to a smaller size of the acceleration/emission region \cite{2011ApJ...742L..32M}. Therefore, the spectral curvature of EED is expected to be proportional to the size of blazar region. As the $\gamma$-rays from BL Lacs are believed to be SSC emission of the same non-thermal electron population emitting low energy synchrotron emission, a smaller size would enhance the synchrotron seed photons energy density $u_s= L_s/4\pi R^2 c \delta^4$ in the jet, that would lead to higher Compton dominance $CD=u_s/u_B$, i.e, a higher relative $\gamma$-ray power. This implies that among the BL Lacs with same a synchrotron luminosity $L_s$, the source with relatively high $\gamma$-ray dominance may have relatively compact emission region.

\cite{2010ApJ...716...30A} provided broadband simultaneous or quasi-simultaneous spectral data from radio through $\gamma$-rays, of 48 LAT Bright AGNs, within the first 3 months operation of Fermi-LAT, from 2008 August 4 to October 31. The broadband spectral data are derived from many ground-based and space-based observatories, including Swift (UVOT, XRT, BAT), Effelsberg, OVRO, RATAN, GASPWEBT, Spitzer-MIPS and AGILE. We show an example of SEDs of two BL Lacs (PKS 0048-09 and S5 0716+714), which are presented in Figure \ref{fig:theory}. These sources have similar synchrotron luminosity and peak frequency but different $\gamma$-ray power, i.e., the Compton dominance. Their quasi-simultaneous broadband SEDs are compiled from \cite{2010ApJ...716...30A} shown as large and small black squares and magenta circles for PKS 0048-09 and S5 0716+714, respectively. The S5 0716+714 included 3 observations at optical and X-ray bands during those three months \cite{2010ApJ...716...30A}. We use the average value of fluxes for this BL Lac object.

We use standard one-zone SSC model \cite{2013ApJ...765..122Y, 2014ApJ...788..104Z, 2014ApJS..215....5K, 2014Natur.515..376G, 2017ApJ...842..129C} to fit their SEDs. One-zone model is widely used in blazar SED modeling \cite{2014Natur.515..376G, 2017ApJ...842..129C}. Since the coordinated variability in different wavelength bands is often seen (although not always) in blazars, a one-zone model assumes that bulk of the non-thermal emission are produced from a "one-zone" region, mostly assumed to be a spherical blob with radius $R$ embedded in a homogeneous but tangled magnetic field $B$ and filled with emitting electron population. Since the considered emitting region is always compact, its synchrotron self-absorption frequency is always large, therefore the model cannot account for the radio flux at observed frequencies smaller than a few hundreds GHz, which are produced by the superposition of several larger components \cite{1979ApJ...232...34B, 1981ApJ...243..700K}. The emitting relativistic blob moves with angle $\theta$ with the line of sight, yielding a Doppler beaming factor $\delta$. In order to keep accordance with log-parabolic shape of synchrotron SED bump, the emitting particles are assumed to be leptons with energy distribution following a log-parabolic function,
\begin{equation}
N(\gamma)=N_{0}\left(\frac{\gamma}{\gamma_{0}}\right)^{-3}10^{-r \log
\left(\gamma/\gamma_{0}\right)^2},
\end{equation}
where $r$ measures the curvature of number distribution of electron energy \cite{2017ApJ...842..129C}. As discussed above, the curvature may be proportional to the size of emission region in the SSC model. Therefore, in our calculation, the curvature parameter ($\beta$) is assumed to be proportional to the radius of the emission sphere, i.e., $r\propto R$. The calculated SEDs are also presented in the Figure \ref{fig:theory}. It seems interesting that both SEDs can be well fitted with same jet parameters except for the radii of emission sphere $R_{0048}=2.8\times10^{16}$ cm and $R_{0716}=5.0\times10^{16}$ cm for PKS 0048-09 and S5 0716+714 (and therefore the curvature $r_{0048}=0.5$ and $r_{0716}=\left(R_{0716}/R_{0048}\right)r_{0048}=0.89$) respectively. The other constant jet parameters are the magnetic field strength $B=0.15$ G, the Doppler factor $\delta=24$, the peak electron energy $\gamma_{0}=8660$ and the normalized total electron numbers $N_{0}=6.8\times10^{45}$. This modeling example illustrates that a higher $\gamma$-ray luminosity intrinsically accompanies a smaller curvature, which may be related to a smaller size of emission region and the processes governing the jet micro-physics, i.e., particle acceleration and cooling. This is further consistent with the fact that bright $\gamma$-ray blazars are more (rapidly) variable than $\gamma$-ray weak blazars from studying $\gamma$-ray and optical variability of a sample blazars \cite{2013ApJ...767..103V, 2014MNRAS.439..690H, 2016MNRAS.455L...1W}, possibly due to smaller size of emission region and/or larger Doppler beaming factor in bright sources. Figure \ref{fig:theory} shows that among the BL Lacs with a given synchrotron peak luminosity and frequency, the object with the highest CD might have smaller curvature (i.e., hard spectrum). A high CD coupled with a lower curvature favor the detection of a blazar by Fermi. Therefore, we argue that CD and curvature both are important parameters for Fermi detection.

\section{Summary} \label{sec:summary}
We investigate why are some BL Lacs detected having $\gamma$-ray emission by Fermi but others not. We find that the SED curvature and $\gamma$-ray dominance of BL Lacs might be intrinsically related. We select 170 BL Lacs with synchrotron SED curvature $b$ and Doppler factor $\delta$ reported in literature, and divide them into FBLs and NFBLs. We find that FBLs have smaller curvature than that of NFBLs even after getting rid of the beaming effect. We show an example that two Fermi BL Lac objects PKS 0048-09 and S5 0716+714 have similar synchrotron peak frequency and luminosity but different Compton dominance. The PKS 0048-09 shows relatively high $\gamma$-ray dominance as compared to S5 0716+714. Within a one-zone SSC model, we find their quasi-simultaneous SEDs can be well fitted by same physical jet parameters except for the size of emission region and EED curvature (assuming curvature being proportional to the size). The PKS 0048-09 manifest relatively smaller source size and a smaller curvature as compared to S5 0716+714. These results imply that the difference in curvature might be related to $\gamma$-ray dominance and may account for the intrinsic discrepancy between FBLs and NFBLs. A broader SED with compact jet size demands an efficient stochastic acceleration. As the emitting particles have less time to spend in a compact jet, the the acceleration gain or number of acceleration steps should be large. Thus, a more efficient stochastic acceleration in compact jets of FBLs makes the EED relatively broad, yielding a smaller value of curvature parameter. However, a smaller source size, in turn, implies a higher synchrotron photon energy density in the jet, producing powerful $\gamma$-ray emission through SSC process. Therefore at a given synchrotron luminosity, the FBLs may have intrinsically smaller curvature and higher Compton dominance as compared to NFBLs. A study of large sample in future may be necessary to further investigate the relationship of curvature and $\gamma$-ray dominance of blazars.

%\acknowledgments
%We thank the anonymous referees for their valuable comments and suggestions that helped improve the manuscript. This research has made use of the NASA/IPAC Extragalactic Database (NED), which is operated by the Jet Propulsion Laboratory, California Institute of Technology, under contract with the National Aeronautics and Space Administration. Part of this work is based on archival data, software, or online services provided by the Space Science Data Center - ASI. This work is supported by the National Natural Science Foundation of China (grant no: U1831138 and 11873073).

\bibliographystyle{unsrt}  
\bibliography{ref}

\end{document}